\documentclass[aps,prl,twocolumn,showpacs,superscriptaddress,groupedaddress,
amsmath,amssymb,floatfix,nofootinbib]
{revtex4}

\usepackage{graphicx}
\usepackage{dcolumn}
\usepackage{bm}
\usepackage[usenames,dvipsnames]{color}
\usepackage{amsmath,amssymb}
\usepackage{slashed}
\usepackage{comment}
\usepackage{hyperref}

\begin{document}

%    Greek Letters
\let\a=\alpha      \let\b=\beta       \let\c=\chi        \let\d=\delta
\let\e=\varepsilon \let\f=\varphi     \let\g=\gamma      \let\h=\eta
\let\k=\kappa      \let\l=\lambda     \let\m=\mu
\let\o=\omega      \let\r=\varrho     \let\s=\sigma
\let\t=\tau        \let\th=\vartheta  \let\y=\upsilon    \let\x=\xi
\let\z=\zeta       \let\io=\iota      \let\vp=\varpi     \let\ro=\rho
\let\ph=\phi       \let\ep=\epsilon   \let\te=\theta
\let\n=\nu
\let\D=\Delta   \let\F=\Phi    \let\G=\Gamma  \let\L=\Lambda
\let\O=\Omega   \let\P=\Pi     \let\Ps=\Psi   \let\Si=\Sigma
\let\Th=\Theta  \let\X=\Xi     \let\Y=\Upsilon

%    Calligraphic letters

\def\cA{{\cal A}}                \def\cB{{\cal B}}
\def\cC{{\cal C}}                \def\cD{{\cal D}}
\def\cE{{\cal E}}                \def\cF{{\cal F}}
\def\cG{{\cal G}}                \def\cH{{\cal H}}
\def\cI{{\cal I}}                \def\cJ{{\cal J}}
\def\cK{{\cal K}}                \def\cL{{\cal L}}
\def\cM{{\cal M}}                \def\cN{{\cal N}}
\def\cO{{\cal O}}                \def\cP{{\cal P}}
\def\cQ{{\cal Q}}                \def\cR{{\cal R}}
\def\cS{{\cal S}}                \def\cT{{\cal T}}
\def\cU{{\cal U}}                \def\cV{{\cal V}}
\def\cW{{\cal W}}                \def\cX{{\cal X}}
\def\cY{{\cal Y}}                \def\cZ{{\cal Z}}

\newcommand{\Ns}{N\hspace{-4.7mm}\not\hspace{2.7mm}}
\newcommand{\qs}{q\hspace{-3.7mm}\not\hspace{3.4mm}}
\newcommand{\ps}{p\hspace{-3.3mm}\not\hspace{1.2mm}}
\newcommand{\ks}{k\hspace{-3.3mm}\not\hspace{1.2mm}}
\newcommand{\des}{\partial\hspace{-4.mm}\not\hspace{2.5mm}}
\newcommand{\desco}{D\hspace{-4mm}\not\hspace{2mm}}

\def\be{\begin{equation}}
\def\ee{\end{equation}}
\def\bea{\begin{eqnarray}}
\def\eea{\end{eqnarray}}
{\newcommand{\lsim}{\mbox{\raisebox{-.6ex}{~$\stackrel{<}{\sim}$~}}}
{\newcommand{\gsim}{\mbox{\raisebox{-.6ex}{~$\stackrel{>}{\sim}$~}}}
\def\Mpl{M_{\rm {Pl}}}
\def\Mp{M_{\rm {P}}}
\def\gev{{\rm \,Ge\kern-0.125em V}}
\def\tev{{\rm \,Te\kern-0.125em V}}
\def\mev{{\rm \,Me\kern-0.125em V}}
\def\ev{\,{\rm eV}}
\def\Treh{T_{\rm {reh}}}
\def\Tf{T_{\rm {f}}}
\def\TEW{T_{\rm EW}}
\def\treh{t_{\rm {reh}}}
\def\Hreh{H_{\rm {reh}}}
\def\Hmax{H_{\rm {max}}}
\def\Rmax{R_{\rm {max}}}
\def\nG{n_{\tilde G}}
\def\YG{Y_{\tilde G}}
\def\mG{m_{\tilde G}}
\def\mGz{m_{\tilde G0}}
\def\rhoG{\rho_{\tilde G}}
\def\mg{m_{\tilde g}}
\def\greh{g_{*\rm{reh}}}
\def\Sigmatot{\Sigma_{{\rm {tot}}}}

\newcommand{\vev}[1]{\langle #1 \rangle}

\newcommand{\td}{t_d}
\newcommand{\tkin}{t_{kin}}
\newcommand{\tthr}{t_{thr}}
\newcommand{\tf}{t_f}
\newcommand{\Gt}{{\widetilde{G}}}
\newcommand{\gt}{{\tilde{g}}}
\newcommand{\MeV}{\,\textrm{MeV}\,}
\newcommand{\GeV}{\,\textrm{GeV}\,}
\newcommand{\OR}{\,\textrm{or},\qquad}
\newcommand{\calA}{\mathcal{A}}
\newcommand{\half}{\frac{1}{2}}
\newcommand{\paren}[1]{\left( #1 \right)}
\newcommand{\dslash}{\slashed{\partial}}
\def\gev{{\rm \,Ge\kern-0.125em V}}
\def\tev{{\rm \,Te\kern-0.125em V}}

\title{\boldmath 
Gravitino production in a thermal Universe revisited}

\author{Richa Arya}
\email{richaarya@prl.res.in, richa.arya@iitgn.ac.in}
\affiliation{Theoretical Physics Division, Physical Research Laboratory,
Navrangpura, Ahmedabad 380 009, India}
\affiliation{IIT Gandhinagar, Palaj, Gandhinagar 382 355 , India}
\author{Namit Mahajan}
\email{nmahajan@prl.res.in}
\author{Raghavan Rangarajan}
\email{raghavan@prl.res.in}
\affiliation{Theoretical Physics Division, Physical Research Laboratory,
Navrangpura, Ahmedabad 380 009, India}

\date{\today}

\begin{abstract}

We study the production of spin 1/2 gravitinos in a thermal Universe. 
Taking into account supersymmetry breaking due to the finite thermal energy 
density of the Universe, there is a large enhancement in the cross 
section of production of these gravitino states.  We consider gravitinos with zero temperature
masses of 0.1 eV, 1 keV, 100 GeV and 30 TeV as representative of
gauge mediated, gravity mediated and anomaly mediated supersymmetry breaking
scenarios.  We find that the 
abundance of gravitinos produced in the early Universe is very high for
gravitinos of mass 1 keV and 100 GeV.
The gravitino 
abundances can be sufficiently suppressed if the reheat temperature is 
less than 100 GeV and $4\times10^4 \gev$ respectively.  
However such low reheat temperatures will rule out 
many models of baryogenesis including those via leptogenesis.

\end{abstract}


\maketitle

\section{Introduction}

Local supersymmetry, or supergravity, gives us a massless spin 2 particle that one can identify with the graviton, the intermediate boson for gravitational interactions. 
The superpartner of the graviton is the massless gravitino with spin
states $\pm 3/2$. When supersymmetry breaks, the gravitino gains mass and spin
$\pm 1/2$ states via a super-Higgs mechanism.  
The spin $\pm 1/2$ states of the gravitino are often referred to as goldstino modes.

Gravitinos are produced in the early Universe  either in the radiation dominated Universe
after reheating 
\cite{
Nanopoulos:1983up,
Krauss:1983ik,
Falomkin:1984eu,Khlopov:1984pf,
Ellis:1984eq,Juszkiewicz:1985gg,Ellis:1984er,
Khlopov:1993ye,Moroi:1993mb,
Kawasaki:1994af,Bolz:2000fu,
Cyburt:2002uv,
giudiceetal1,
kkm,Kohri:2005wn,
Pradler:2006qh,
pradlersteffen2},
or 
during standard (perturbative) reheating 
\cite{giudiceetal1,kkm,pradlersteffen2,Rangarajan:2006xg,Rangarajan:2008zb,Rychkov:2007uq},
by the scattering of thermalised inflaton decay products. Gravitinos can also be produced during preheating 
\cite{giudiceetal1,maroto.00,kallosh.00,tsujikawa.00,
Nilles:2001fg,
nilles&olive.01,Greene:2002ku,podolsky,nilles.01,Fujisaki:1995ua}
or via direct
inflaton decay \cite{Kawasaki:2006gs,Kawasaki:2006mb,
Nakayama:2012hy,Endo:2007sz}, or during and after inflation in 
warm inflation scenarios
 \cite{Taylor:2000jw,Bartrum:2012tg}.
As argued in Ref. \cite{Bolz:2000fu}, the 
gravitino production rate in supersymmetric QCD via scattering at high temperature is proportional to 
\be
\frac{1}{\Mp^2}\left(1+ \frac{m_\gt^2}{3 m_\Gt^2}\right)
\label{bolz_sigma}
\ee
where $\Mp
\simeq
2.4\times 10^{18}$ GeV is the reduced Planck mass,  $m_\gt$ is the explicit supersymmetry breaking gluino mass and $m_\Gt$ is the gravitino mass.  
The first term within the parentheses is associated with
spin $3/2$ gravitino production
while the second term is associated with spin $1/2$ gravitino production.

Excessive abundance of gravitinos creates cosmological problems.  A very light 
($ m_{\tilde{G}}\ll 1$ MeV) and stable gravitino acts as an additional relativistic degree of freedom during primordial nucleosynthesis and can affect the expansion rate and thereby the light nuclear abundances
(depending on its contribution to the effective relativistic degrees of freedom). For a stable gravitino of mass 
greater than 1 keV,
its energy density today turns out to be  higher than the critical density and it can overclose the Universe. 
A gravitino of mass between $100 \ {\rm GeV}\lsim m_{\tilde{G}}\lsim 10$ TeV decays into 
energetic particles after nucleosynthesis which dissociate light nuclei created during primordial
nucleosynthesis.
The extent of impact of the gravitinos on the cosmology of our Universe depends directly on its abundance.

The initial calculation of the gravitino abundance done in Refs. \cite{Nanopoulos:1983up,Khlopov:1984pf,Ellis:1984eq,Krauss:1983ik,Falomkin:1984eu} considered gravitino production in the radiation dominated
Universe after reheating for spin $3/2$ states. 
It was found that the abundance $Y_\Gt \equiv n_\Gt/s$, where
$n_\Gt$ is the gravitino number density and $s$ is the entropy density, is proportional to the reheat temperature $\Treh$. This then gave
an upper bound on the reheat temperature. Many subsequent estimates of the gravitino abundance created in the radiation dominated Universe after reheating considered different channels for gravitino decay as a function of the gravitino mass and obtained associated upper bounds on the 
abundance or reheat temperature \cite{kkm,Kohri:2005wn}. 
Again, these works also considered only spin $3/2$ states.   
As mentioned in 
Ref. \cite{kkm}
considering only the spin 3/2 states gives a conservative
estimate of the gravitino abundance.  Furthermore, 
for gravity mediated supersymmetry breaking the second
term in the parantheses in Eq. (\ref{bolz_sigma}) is of $O(1)$ and so the abundance obtained is of the right order in this case.

In the present work we study the production of spin 1/2 gravitinos in the radiation dominated Universe after reheating. Spin 1/2 gravitinos are associated with goldstino modes and, as we argue below,
their production cross section should not be Planck mass suppressed but instead 
suppressed by the supersymmetry breaking scale in the hidden sector.
We then argue that the finite energy density of a thermal Universe also breaks supersymmetry, and
in scenarios where 
interactions in the thermal bath are mediated by light particles the finite energy density
affects both the fermion-boson mass squared splitting and the gravitino mass, and thus
the goldstino production cross 
section.

In the standard picture of hidden sector supersymmetry breaking we have a hidden sector with fields $[H]$, a visible sector with fields $[V]$ and
a messenger sector that mediates the supersymmetry breaking with fields $[X]$ with mass $M_X$. Supersymmetry breaks in the hidden sector, say, by F-term breaking with 
$\langle F_H\rangle = f_H$. The soft supersymetry breaking mass in the visible sector that gets generated due to the interaction between the visible and hidden sectors
mediated by the messenger sector is
\bea
m_{\rm soft} \sim \frac{1}{M_X} \langle F_H\rangle \label{zeroTmsoft}\,.
\eea
For phenomenological reasons, we require $m_{\rm soft} \sim 
100 \ \gev$. 
This then, depending upon the mediation mechanism, sets the scale for $\langle F_H\rangle$.

The goldstino coupling to matter will be proportional to the mass squared splitting between particles and their superpartners. 
From Eq. (\ref{bolz_sigma}) the production rate for spin 1/2 gravitino states is
\bea
\Gamma_{s}&\sim& \frac{1}{\Mp^2}\frac{m_\gt^2}{m_\Gt^2}
\label{Gamma_half1}\\
& \sim&\frac{1}{\Mp^2}\frac{m_{\rm soft}^2}{m_\Gt^2}
\label{Gamma_half2}
\\
&\sim&\frac{1}{\Mp^2}\frac{m_{\rm soft}^2}{(M_S^2/\Mp)^2}\cr
&\sim&\frac{1}{M_S^2} \frac{m_{\rm soft}^2}{M_S^2}\,
\eea
where $m_{\rm soft}^2$ is the  mass squared splitting between superpartners while
$M_S=\sqrt{\langle F_H\rangle}$ is the scale of supersymmetry breaking.
It can be further noted that
the goldstino production rate above is not Planck mass suppressed but suppressed by the supersymmetry
breaking scale $M_S$.
The production rate goes to
zero in the supersymmetric case.

It is known that supersymmetry is broken by non-zero temperature $T$.
It
has also been shown 
that the effect of  the non-zero temperature is to split the boson and fermion masses, with the splitting $m_{{\rm soft},T}^{2} \sim g^2 T^2$, where $g$ is a generic coupling constant,
which we refer to as soft mass generation due to finite temperature effects. 
That supersymmetry is broken by finite temperature effects can also be seen by the following
argument: 
in
the high temperature limit
we know that the theory gets dimensionally reduced to a lower dimension. All the fermion Matsubara
modes (recall that there is no $n=0$ mode for fermions) become heavy while for bosons all modes become heavy except the $n=0$ mode. Thus, the
low energy effective theory will only contain a bosonic field and no fermionic field, and therefore 
supersymmetry is broken.
Moreover in Refs. \cite{Derendinger:1998zj,Lucchesi:1998eb}
it has been shown, by invoking a thermal superspace 
approach and applying it to systems of thermal fields, that 
supersymmetry is explicitly broken at finite temperature and that
the thermal action is not invariant under thermal supersymmetry.
Unlike in other works, we consider supersymmetry breaking due to
the finite energy density of the radiation dominated Universe, $\rho=(\pi^2/30)g_*T^4$,
where $g_*\sim228.75$ is the effective
number of relativistic degrees of freedom.

Let us now 
consider the effect of having a finite energy density, $\rho \sim T^4$.
Consider three chiral superfields, $S,\, \Phi$ and $Y$ having a coupling 
$\lambda S\Phi Y$, where 
$\lambda$
is the coupling, $\Phi$ belongs to $[V]$
and $S$  
could belong to $[V]$, $[H]$ or $[X]$.  We 
assume that $S$ contributes to the radiation energy density of the Universe. 
In the superfield language, the four point amplitude $\Phi^{\dag}\Phi S^{\dag}S$ reads
(we have employed the off-diagonal component of the GRS propagator for $Y$ and retained the
external fields) 
\begin{eqnarray}
 {\mathcal{A}}^{(4)} =& \vert\lambda\vert^2
 \Phi^{\dag}S^{\dag}\,
& \left[\int d^4x d^4x'd^4\theta d^4\theta' \right.  \nonumber\\
 &&\left. \frac{1}{\Box^2 - m_Y^2}
  \delta(z-z')\right] 
  \Phi S
\end{eqnarray}
where $z=(x,\theta,\bar\theta)$ and $\delta(z-z') = \delta^4(x-x')\delta^2(\theta-\theta')\delta^2(\bar{\theta}-\bar{\theta'})$. 
In a thermal bath, the typical $Y$ momentum  
$q \sim Q$,  $i.e.$ there is a distribution peaked at $Q\sim T$ (or $Q\sim \sqrt{T m_Y})$, 
for $T\gg m_Y$ (for $T\ll m_Y)$, which for simplicity we
take to be $\delta(q-Q)$. 
For $T\ll m_Y$ we then get 
${\mathcal{A}}^{(4)}\sim \vert\lambda\vert^2/m_Y^2 \,(\Phi^{\dag}S^{\dag}\Phi S)$, 
while for $T\gg m_Y$, we obtain
${\mathcal{A}}^{(4)}\sim \vert\lambda\vert^2/T^2 (\,\Phi^{\dag}S^{\dag}\Phi S)$.
This is equivalent to having a term in the effective Lagrangian as
\begin{equation}
{\cal L}_{\rm eff}
= 
\frac{\vert\lambda\vert^2}{m_Y^2,T^2}\,\Phi^{\dag}S^{\dag}\Phi S 
\,.
\label{Leff}
\end{equation}
Now we expand  
${\cal L}_{\rm eff}$
in all powers of $\theta$ and $\bar\theta$.  The 
relevant
$\theta\theta\bar\theta\bar\theta$ term for us will be, 
in a thermal environment,
\begin{equation}
\frac{\vert\lambda\vert^2}{m_Y^2,T^2}\,
\langle S^\dagger S\vert_{\theta\theta\bar\theta\bar\theta} \rangle_{\rm thermal}
\phi^\dagger\phi\,,
\end{equation}
where $\phi$ is the scalar component of $\Phi$. 
The term
$\langle S^\dagger S\vert_{\theta\theta\bar\theta\bar\theta} \rangle_{\rm thermal}$ above is
$\langle i \partial_m\bar\psi_S\bar\sigma^m\psi_S + s^*\Box s + F_S^*F_S \rangle_{\rm thermal}
$.  The first two terms can be identified with the kinetic energy terms in the Lagrangian for
the fermionic and scalar components of the superfield $S$.
$|F_S|^2=|\partial W/\partial s|^2 =V(s)$ 
and one may consider, say, a 
quartic potential for $s$.  Then $\langle S^\dagger S\vert_{\theta\theta\bar\theta\bar\theta} \rangle_{\rm thermal} \sim T^4$.
Therefore we obtain
\begin{equation}
m_{\rm soft}^2 \sim\vert\lambda\vert^2 \frac{T^4}{m_Y^2} \,\,\,\,\,{\rm for} \,\,T\ll m_Y
\label{smallTmsoft1}
\end{equation} 
and
\begin{equation}
m_{\rm soft}^2 \sim\vert\lambda\vert^2 T^2 \,\,\,\,\,\,\, {\rm for}\, \,T\gg m_Y.
\label{largeTmsoft1}
\end{equation}
(The above example can be suitably extended to vector superfields. Also, there will be, in general, more than one such contribution to the soft masses.
We have chosen the simplest one to bring out the essence of the 
argument.)

Comparing Eq. (\ref{smallTmsoft1}) and Eq. (\ref{zeroTmsoft}), the two forms are quite similar,
and when $S$ belongs to, say, $[H]$, and the amplitude is mediated by a heavy field  
it is quite natural to assume
$m_Y \sim M_X$. 
In such a case, it appears that the finite temperature effects essentially look like an additional contribution to the F-term breaking.
For $f_H \gg T^2$
one does not have any large temperature dependent contribution to the soft breaking masses, and therefore no enhanced contribution to goldstino couplings. A similar conclusion is reached if $S$ is one of the visible sector fields and the interaction
between $S$ and $\Phi$ is mediated by a third visible sector field which is very massive.
Refs. \cite{Leigh:1995jw,Ellis:1995mr,Rychkov:2007uq} also argue that finite temperature effects will not lead to enhanced goldstino production as originally argued in Ref. \cite{Fischler:1994uw}.

Now consider the case when the four point superfield amplitude is mediated by a light $Y$ 
superfield. This is quite reasonable to expect  
since in thermal equilibrium
different fields in the visible sector, for example, $\Phi$ and $S$, can interact via, say, gauge/Yukawa interactions such that the mediator is a 
massless/light field. 
In this case, $i.e.$ when the amplitude is mediated by the light field, the soft supersymmetry breaking mass in Eq. (\ref{largeTmsoft1}) 
contributes to the scale of the mass splitting between the
bosonic and fermionic partners.

From the above discussion 
we observe that depending up on the mass scale of the field that mediates the four point amplitude,
the finite temperature contribution to the soft breaking mass takes the form
\begin{equation}
m_{\rm soft} ^2
\sim \frac{1}{M_X^2}(f_H^2 + \delta^2 T^4),
\,\,\, m_Y\sim M_X \gg T
\label{msoft^2largeTother}
\end{equation} 
where $\delta$ is some parameter,
or 
\begin{equation}
m_{\rm soft}^2 \sim \frac{1}{M_X^2}f_H^2 + \delta T^2, \,\,\, m_Y \ll T\,.
\label{msoft^2largeT}
\end{equation}

Below we shall assume that $S$ is one of the visible sector fields and 
there is naturally a massless/light field that mediates the four point amplitude,
and therefore what is relevant is Eq. (\ref{msoft^2largeT}).
We would like to emphasize that 
this is exactly where we differ from the usual treatment of finite temperature effects in the context of
gravitinos.
Then in Eq. (\ref{Gamma_half1})
\be
m_\gt^2\rightarrow m_\gt^2-m_g^2\sim \delta_3 T^2 +m_0^2\,,
\ee 
where $m_g$ is the gluon mass, $\delta_3$ is some parameter 
and $m_0\sim 100\gev$ represents the zero temperature mass splitting, 
while 
\bea
m_\Gt
\sim
 \sqrt\rho/(\sqrt3 M_P)+m_{\Gt 0} \nonumber \\
=\delta' T^2/(\sqrt3 M_P)+m_{\Gt 0} 
\eea
where $m_{\Gt 0}$ is the zero temperature 
gravitino mass (which depends on the supersymmetry breaking mechanism relevant at low temperatures), and 
$\delta'$ is another parameter.
Then the factor in the scattering rate in Eq. \eqref{Gamma_half2}
\be
\gamma_3\equiv\frac{m_{\rm soft}^2}{3m_\Gt^2}=
\frac{\delta_3 T^2+m_0^2}{3[\delta' T^2/(\sqrt3 M_P)+m_{\Gt0}]^2}.
\label{susy1}
\ee
When $\delta_3 T^2\gg m_0^2\ {\rm and} \ \delta' T^2 /(\sqrt3 M_P)\gg m_{\Gt0}$
\be
\gamma_3\approx \frac{\delta_3 T^2}{(\delta' T^2/M_P)^2} 
=\frac{\delta_3}{\delta^{'2}} \frac{M_P^2}{T^2}
\label{susy2}
\ee
which can be much larger than 1.  
This
can be much larger than
the zero temperature limit $m_0^2/(3m_{\Gt0}^2)$ in $\Gamma_s$,
and therefore necessitates a fresh look at the calculation of the gravitino
abundance.  
(It may be noted, however, that our final results  will depend only on
the zero temperature form of the gravitino mass.)

Unlike in the standard calculations of the gravitino abundance, in our scenario gravitinos
will be in thermal equilibrium in the early Universe because of the enhanced scattering
rates.  The gravitinos decouple when they are relativistic, 
and hence they have a large abundance as a hot relic. 
Below we shall consider gravitinos with zero temperature masses 
of 0.1 eV, 1 keV, 100 GeV, 30 TeV. Typically one can obtains such masses in gauge 
mediated (0.1 eV, 1 keV), gravity mediated and anomaly mediated supersymmetry 
breaking scenarios respectively. We find that  the very light gravitinos ($m_{\Gt0}=0.1$ eV) 
have $\Omega_{\tilde G}\ll 1$, and hence will not overclose the Universe. The very heavy 
gravitinos ($m_{\Gt0}=30$ TeV) have a very large abundance but have a short lifetime and 
decay before nucleosynthesis and do no greatly alter the cosmology of the Universe. 
However, the gravitinos with mass $\sim$ 1 keV have $\Omega_{\tilde G} \sim 1$, so can 
affect the cosmology of our Universe. We further find that the abundance of $m_{\Gt0}=100$ 
GeV gravitinos is orders of magnitude higher than the 
cosmological upper bound.  The abundances can be suppressed by considering a low
reheat temperature less than  100 GeV and $4 \times 10^4 \gev$ for the 1 keV and
100 GeV gravitinos respectively.  Such low reheat temperatures will rule
out models of high scale baryogenesis including those via leptogenesis.  Very low scale
baryogenesis scenarios, and electroweak baryogenesis
and low scale leptogenesis models respectively will then be the preferred mechanisms for
generating the matter-antimatter asymmetry of the Universe.

In an earlier work we had studied the production of spin 1/2 gravitinos in the presence
of supersymmetric flat directions which give mass to some gauge bosons, gauginos, and 
sfermions and had found that there is resonant production of (spin 1/2)
gravitinos leading
to an extremely large abundance which is orders of magnitude larger than the cosmological bound \cite{Mahajan:2013wma}.
Below we do not consider the presence of supersymmetric flat directions.
\footnote
{If the supersymmetric flat direction gives mass to {\it all} gauge bosons it delays thermalization
of the inflaton decay products leading to suppressed gravitino production, as discussed in Refs.
\cite{Allahverdi:2005fq,Allahverdi:2005mz,Rangarajan:2012wy}.
In these works, the thermal
energy density contribution to supersymmetry breaking was not included.
}

\section
{Gravitino production}

As mentioned earlier, gravitinos are produced by the scattering of the thermalised decay products
of the inflaton. The different processes that produce gravitinos are listed in Table 1 in  
Refs. \cite{Ellis:1984eq,Kawasaki:1994af} 
and Table 4.3 of Ref. \cite{Moroi:1995fs}. These processes include, for example, 
$g {\tilde g} \rightarrow g {\tilde G},
{\tilde q} g \rightarrow q  {\tilde G},
q {\bar q} \rightarrow {\tilde g} {\tilde G}$, etc.
Besides these scattering processes there are also annihilation processes, 
$\tilde{G} \tilde{G} \rightarrow \gamma \gamma$ and $\tilde{G} \tilde{G} \rightarrow f \bar{f}$, 
as we shall see below.

The thermally averaged cross section $\langle\Sigmatot|v|\rangle$
for the scattering processes in Refs.~\cite{Ellis:1984eq,Kawasaki:1994af}
is given by~\cite{Pradler:2006qh}
\be
\langle\Sigmatot|v|\rangle \equiv
\frac{\alpha}{\Mp^2}\nonumber\\
\ee
\bea
= \frac{1}{\Mp^2} \frac{3\pi}{16\zeta (3)}
\sum_{i=1}^3
\left[ 1+
\frac{m_i^2}{3\mG^2}
\right]
c_i \,g_i^2\, \ln\left(\frac{k_i}{g_i}\right)
\label{sigma-total}
\eea
where $i=1,2,3$ refers to the three gauge groups $U(1)_Y, SU(2)_L$ and 
$SU(3)_c$ respectively,\ $g_i(T)$ are the gauge coupling constants (evaluated at the
most relevant temperature), and
$c_{1,2,3}=11, 27,72$ and $k_{1,2,3}=1.266, 1.312, 1.271$
are constants associated with the gauge groups
 (see Table 1 of Ref.~\cite{pradlersteffen2}). The above
expression includes corrections to the cross section
for gravitino production obtained earlier
in 
Refs.~\cite{Bolz:2000fu} and \cite{kkm}. 
We have also replaced the gaugino mass squared in the original expression with 
$m_i^2=\delta_i T^2 +m_0^2=m_{\rm soft} ^2$ representing the difference in gaugino and gauge boson masses squared.
(In the current analysis we have ignored the possibility of a Breit-Wigner resonance associated with incoming particles of energy
$\sim T$ and the intermediate supersymmetric particle having a mass of $O(T)$.  We hope to return to this
issue in a future publication.)

The rate of production of gravitinos for the processes listed in  Table 1 of \cite{Ellis:1984eq,Kawasaki:1994af} is given by
\[\Gamma_{s}=n\langle\Sigmatot|v|\rangle \]
where $n=({\zeta(3)}/{\pi^2})T^3$ is the the number density of the scatterers, and 
the Riemann zeta function
$\zeta(3)=1.2020...$.
Then, taking all $m_i=m_{3}$, we get
\begin{equation}
\Gamma_{s}= \frac{3T^3}{16\pi M_P^2}\left(1+ \frac{m_3^2}{3 m_{\tilde{G}^2}}\right)\sum_{i=1}^3 c_i g_i^2 \ln\left(\frac{k_i}{g_i}\right).
\label{Gammasfinal}
\end{equation}

We now consider different cases for $\gamma_3$ defined in Eq. (\ref{susy1}).
 \begin{itemize}
\item \textbf{Region I :} $\delta_3 T^2 > m_0^2$ and $\delta' T^2/(\sqrt{3}M_P)>m_{\tilde{G0}}$. Then
 \[\gamma_3 
 %=
 \approx 
 \frac{\delta_3 M_P^2}{\delta'^2 T^2}.\]
  
 \item \textbf{Region II :}  $\delta_3 T^2>m_0^2$ 
and $\delta' T^2/(\sqrt{3}M_P)< m_{\tilde{G0}}$. Then
 \[\gamma_3 
 %= 
 \approx
 \frac{\delta_3 T^2 }{3m_{\tilde{G0}}^2}. \]

 \item \textbf{Region III :} $\delta_3 T^2<m_0^2$ and $\delta' T^2/(\sqrt{3}M_P)<m_{\tilde{G0}}$. Then 
\[\gamma_3 
%= 
\approx
\frac{ m_0^2}{3m_{\tilde{G0}}^2}.\]
\end{itemize}
We shall take $\delta_3, \delta'\sim 0.1$.  
As we shall see below, 
scattering processes 
will maintain the gravitinos in thermal equilibrium
in our scenario till they freeze out.  Thereafter annihilation processes such as
$\tilde{G}\tilde{G} \rightarrow \gamma \gamma$ and $\tilde{G} \tilde{G} \rightarrow f \bar{f}$ 
become relevant.

For the process $\tilde{G} \tilde{G} \rightarrow \gamma \gamma$, if $\sqrt{s}\gg m_{\tilde{\gamma}}$, 
where $\sqrt s\sim T$ is the centre of mass energy, then the annihilation cross section is given by \cite{gherghetta:1997}
\begin{equation}
\sigma_{\tilde{G} \tilde{G} \rightarrow \gamma\gamma}=\frac{1}{1728\pi}\frac{\kappa^4}{m_{\tilde{G}}^4}m_{\tilde{\gamma}}^4 s
\end{equation}
where $\kappa=\frac{1}{M_{Pl}}$ \rm{and}\ $M_{Pl}=1.2\times 10^{19}$ GeV is the 
Planck mass.
If $\sqrt{s}\ll m_{\tilde{\gamma}}$ then the annihilation cross section is
\begin{equation} 
\sigma_{\tilde{G} \tilde{G} \rightarrow \gamma \gamma}=\frac{1}{576\pi}\frac{\kappa^4}{m_{\tilde{G}}^4}m_{\tilde{\gamma}}^2 s^2=\sigma_{\gamma\gamma}s^2\,,
\label{annihilg-low-s}
\end{equation}
where
$\sigma_{\gamma\gamma}\equiv\frac{1}{576\pi}\frac{\kappa^4}{m_{\tilde{G}}^4}m_{\tilde{\gamma}}^2 $.  For the process
$\tilde{G} \tilde{G} \rightarrow f \bar{f}$, if $\sqrt{s}\gg m_{\tilde{f}}$ then from Ref. \cite{gherghetta:1997}
\begin{equation}
\sigma_{\tilde{G} \tilde{G} \rightarrow f \bar{f}}=\frac{1}{180\pi}\frac{\kappa^4}{m_{\tilde{G}}^4}m_{\tilde{f}}^4 s\,,
\end{equation}
and if $\sqrt{s}\ll m_{\tilde{f}}$ then
\begin{equation}
\sigma_{\tilde{G} \tilde{G} \rightarrow f \bar{f}}=\frac{1}{180\pi}\frac{\kappa^4}{m_{\tilde{G}}^4}s^3=\sigma_{f \bar{f}}s^3\,,
\label{annihilf-low-s}
\end{equation}
where
$\sigma_{f \bar{f}} \equiv \frac{1}{180\pi}\frac{\kappa^4}{m_{\tilde{G}}^4}. $
The total annihilation cross section is given by
\[\sigma_A=\sigma_{\gamma \gamma}s^2+\sum_f\sigma_{f \bar{f}} s^3\]
where all possible fermion pairs in the final state have been summed over. 
The dominant contribution to the total cross section 
for $\sqrt{s}\ll m_{\tilde\gamma}$, as can be seen from Eqs. \eqref{annihilg-low-s}
and \eqref{annihilf-low-s},
is from $\sigma_{\tilde{G} \tilde{G} \rightarrow \gamma \gamma}$, and the thermally averaged cross section times velocity is approximately given by \cite{gherghetta:1997}
\[\langle\sigma v_{\rm Moller}\rangle=1800\frac{\zeta(5)^2}{\zeta(3)^2}\sigma_{\gamma\gamma}T^4\]
where $\zeta(5)=1.0369...$ The  annihilation rate for gravitinos for 
$\sqrt{s}\ll m_{\tilde\gamma}$
is given by
\[\Gamma_A=n_{\tilde{G}}\langle\sigma_Av_{\rm Moller}\rangle \]
where $n_{\tilde{G}}=\frac{3\zeta(3)}{2\pi^2}T^3. $ Then
\begin{equation}
\Gamma_A=\frac{75}{16\pi^3}\frac{\zeta(5)^2}{\zeta(3)}
\frac{\kappa^4}{\mG^4}m_{\tilde{\gamma}}^2T^7 \simeq 0.135 
\frac{\kappa^4}{\mG^4}m_{\tilde{\gamma}}^2T^7.
\end{equation}
For $\sqrt{s}\ll m_{\tilde\gamma}\sim \delta_1 T^2 + m_0^2 $, $T\ll m_0$
and $\delta' T^2/(\sqrt{3}M_P)\ll m_{\tilde{G0}}$ and so 
\begin{equation}
\Gamma_A= 0.135 
\frac{\kappa^4}{\mGz^4}m_{0}^2T^7.
\end{equation}
For higher temperatures we use the appropriate expressions for the annihilation rate.

\section{Calculation of the gravitino freeze out temperature}

In the standard scenario of gravitino production the rate for production is 
small compared to the Hubble parameter $H$.  
Therefore one does not produce very many gravitinos.  The small
number density of gravitinos then
implies that the inverse scattering process is also suppressed.  
However, in our scenario, because of the enhanced gravitino production rate,
$\Gamma_{s}>H$.  Moreover, the inverse process 
will  be unsuppressed because
of the large gravitino abundance.
Therefore 
the gravitinos 
maintain a thermal
distribution, till their interactions freeze out. (We can use the expression for $\Gamma_{s}$ till $T\sim {\rm max}\ ( \mG,m_{\tilde g})$ since the cross section in Eq. (\ref{sigma-total})
presumes the particles are relativistic.)

The freeze out condition is $ \Gamma_{s}(T_f)= H(T_f)=5 \frac{T_f^2}{M_P} $.  
We also consider the gravitino annihilation processes discussed above. 
For each (zero temperature) gravitino mass we consider each of the three 
regions discussed in 
the previous section.

\subsection{Very light gravitino}
 $m_{\tilde{G0}}=0.1\ \rm{eV}$\\
\\
\textit{Region I:}\  $T> 300\ \rm{GeV}$ and $T>6.5\times 10^4\ \rm{GeV}.$
We find that
$\Gamma_{s}> \Gamma_{A} $ and also $\Gamma_{s} >H$. Hence, gravitinos are in thermal equilibrium in this domain and maintain a thermal abundance with $n_{\tilde G}=\frac{3\zeta(3)}{2\pi^2}T^3$.\\
\\
\textit{Region II:}\  $T>300\ \rm{GeV}$ and $T<6.5\times10^4\ \rm{GeV}. $
We find that
$\Gamma_{s} > \Gamma_{A}$
and also $\Gamma_{s} >H$. Hence, gravitinos are in thermal equilibrium in this domain and maintain a thermal abundance.\\
\\
\textit{Region III:}\ $T<300\ \rm{GeV}$. 
 Till $T\sim m_0=100\gev$, $\Gamma_s>\Gamma_A$ and $\Gamma_s>H$. 
 For $T<100\gev$, scattering processes given in Table 1 in Ref. \cite{Ellis:1984eq,Kawasaki:1994af} are kinematically forbidden
 and
 $\Gamma_{\tilde{G}\tilde{G} \rightarrow \gamma \gamma}$ is the relevant process. 
 However we find that
$\Gamma_{\tilde{G}\tilde{G} \rightarrow \gamma\gamma}<H$. \\

Therefore the freeze out temperature  $T_f= 100\gev$.
The abundance of gravitinos at freeze out
is given by$$Y_{\tilde Gf}=Y_{\tilde G}(T_f)=\frac{n(T_f)}{s(T_f)}=\frac{3 \zeta(3)}{2 \pi^2}\frac{45}{2\pi^2g_{*s}(T_f)}.$$
For the MSSM particle content, $g_{*s}\sim 228.75.$
Then 
$$Y_{\tilde Gf}= 1.8\times 10^{-3}.$$
The lifetime of the gravitinos is given by \cite{weinberg}$$t=\frac{M_P^2}{m_{\tilde G}^3}=1.2\times 10^{35} \ \rm{yr}$$
which is much larger than the age of Universe.
 The density parameter of thermally produced gravitinos is given by
\begin{equation}
\Omega_{\tilde G}=\rho_{\tilde G}/\rho_c=m_{\tilde G} Y_{\tilde Gf}s(T_0) / \rho_c.
\label{OmegaG}
\end{equation}
 Taking $\rho_c/s(T_0)= 1.95\times 10^{-9}$ GeV, we get  $\Omega_{\tilde G} \approx 0.92\ \times 10^{-4}$, which impies that the gravitinos will not overclose the Universe.
 Furthermore,
there is no constraint from primordial nucleosynthesis because the effective number of (nearly)
massless neutrino flavors over and above the Standard Model value,
$\Delta N_\nu = [g_*(1\mev)/g_*(T_f)]^{4/3}$ will be 0.02, which is less than
the current upper bound of 0.4 from Planck 2015 \cite{Agashe:2014kda}. 
However the Cosmic Microwave Background Stage 4 experiments hope to probe $\Delta N_\nu$ down to an accuracy
of 0.027 \cite{cmb}.

\subsection{Light gravitino}
 $m_{\tilde{G0}}=1\ \rm{keV}$\\
 \\
\textit{Region I:}\ $T> 300\ \rm{GeV}$ and $T>6.5\times 10^6\ \rm{GeV}.$
We find that
$\Gamma_{s} > \Gamma_{A} $ and also $\Gamma_{s} >H$. Hence, gravitinos are in thermal equilibrium in this domain and maintain a thermal abundance.\\
\\
\textit{Region II:}\ $T>300\ \rm{GeV}$ and $T<6.5\times10^6\ \rm{GeV}. $
We find that
$\Gamma_{s}>\Gamma_A$ in this temperature range. But
$\Gamma_{s}<H$ for $T<600$ GeV. At this temperature, the gravitinos freeze out. 
\\
\\
\textit{Region III:}\ $T<300$ GeV.
Till $T\sim100\gev$, $\Gamma_s > \Gamma_A$ but $\Gamma_s<H$.
For $T<100\gev$, $\Gamma_{\tilde{G} \tilde{G} \rightarrow \gamma \gamma}$ is the 
relevant
process. But
$\Gamma_{\tilde{G} \tilde{G} \rightarrow \gamma \gamma}<H.$
Hence, gravitinos are out of equilibrium in this domain.\\
\\
 Thus the freeze out temperature 
$T_f=600\ \rm{GeV}$
and the abundance of gravitinos  
is given by
\[
Y_{\tilde Gf}
=1.8\times 10^{-3}.
\]
The lifetime of the gravitinos $$t=\frac{M_P^2}{m_{\tilde G}^3}=1.2\times 10^{23} \ \rm{yr}$$
which is much larger than the age of Universe.
 The density parameter in Eq. \eqref{OmegaG} 
 $\Omega_{\tilde G} \approx 0.92$, which is in conflict
 with observations. In order to avoid this, the gravitino mass was bounded to be less than a keV in Refs. \cite{weinberg,pagels}.
 Again,  
$\Delta N_\nu =  0.02$ which is less than the current upper bound.

\subsection{Heavy gravitino}
$m_{\tilde{G0}}=100\ \rm{GeV}.$\\
\\
\textit{Region I:}\ $T> 300\ \rm{GeV}$ and $T>6.5\times 10^{10}\ \rm{GeV}.$
We find that
$\Gamma_{s} > \Gamma_{A}$
and also $\Gamma_{s}>H$. Hence, gravitinos are in thermal equilibrium in this domain and maintain a thermal abundance.\\
\\
\textit{Region II:}\ $T>300\ \rm{GeV}$ and $T<6.5\times10^{10}\ \rm{GeV}. $
We find that
$\Gamma_{s} > \Gamma_{A} $ in this temperature range. But
$\Gamma_{s}< H$ for $T<1.2\times 10^8$ GeV. At this temperature, the gravitinos freeze. 
\\
\\
\textit{Region III:}\ $T<300$ GeV.
Till $T\sim100\gev$, $\Gamma_s > \Gamma_A$ but $\Gamma_s<H$.
Thereafter 
$ \Gamma_{\tilde{G} \tilde{G} \rightarrow \gamma \gamma}$ is the relevant process. But
$\Gamma_{\tilde{G}\tilde{G} \rightarrow \gamma \gamma}<H$ which implies that the gravitinos are out of equilibrium in this domian.\\
\\
 Thus the freeze out temperature 
$T_f=1.2\times 10^8\ \rm{GeV}$
and the abundance of gravitinos at freeze out 
is given by
\[
Y_{\tilde Gf}
=1.8\times 10^{-3}.
\]
This is much larger than the cosmological upper bound on the gravitino abundance of
$10^{-14}$ \cite{Cyburt:2009pg}.

\subsection{Very heavy gravitino}
 $m_{\tilde{G0}}
 =30\ \rm{TeV}$\\
 \\
\textit{Region I:}\ $T> 300\ \rm{GeV}$ and $T>1\times 10^{12}\  \rm{GeV}.$
We find that
$\Gamma_{s} > \Gamma_{A} $
and also $\Gamma_{s} >H$. Hence, gravitinos are in thermal equilibrium in this domain and maintain a thermal abundance.\\
\\
\textit{Region II:}\ $T>300\ \rm{GeV}$ and $T<1\times 10^{12}\ \rm{GeV}. $
We find that
$\Gamma_{s} > \Gamma_{A} $ in this temperature range. But
$\Gamma_{s}< H$ for $T<5.5\times 10^{9}$ GeV. At this temperature, the gravitinos freeze
out. 
\\
\\
\textit{Region III:}\ $T<300$ GeV.
In this domain, the scattering processes are kinematically forbidden
as $T<\mGz$. Hence, $ \Gamma_{\tilde{G} \tilde{G} \rightarrow \gamma\gamma}$ is the relevant process. But
$\Gamma_{\tilde{G}\tilde{G} \rightarrow \gamma \gamma}<H$
which implies that the gravitinos are out of equilibrium in this domain. \\
\\
Thus the freeze out temperature 
$T_f=5.5\times 10^{9}\ \rm{GeV}$
and the abundance of gravitinos 
at freeze out 
is given by
\[
Y_{\tilde Gf}
=1.8\times 10^{-3}.
\]
The lifetime of the gravitinos is $$t=\frac{M_P^2}{m_{\tilde G}^3}=0.1\ \rm{s}$$
which implies that the gravitinos would have decayed before nucleosynthesis and not lead to any cosmological problem.\\

In all the cases considered above, 
$T_f\gsim \mG,\mg$. 
Then
the use of the expression for $\langle \Sigmatot |v|\rangle$ in Eq. \eqref{sigma-total},
which presumes relativistic incoming and outgoing particles, is justified.
Note that in all cases above freeze out occurs in {\it Region II} or {\it III} for which the zero temperature gravitino mass
is the relevant mass.

\section{Discussion}
\label{Discussion1}

For the (zero temperature) gravitino masses  of 0.1 eV, 1 keV, 100 GeV and 30 TeV that we have considered, the 
freeze out temperature is higher than the gravitino mass (including thermal effects).  Then the gravitinos are hot relics
and their abundance 
at freeze out is $Y_{\tilde Gf}\sim 1.8\times 10^{-3}$. 
For the gravitino with zero temperature masses  
of 0.1 eV and 1 keV
there is no constraint from primordial nucleosynthesis because
$\Delta N_\nu = 0.02$ which is less than the current upper bound.
The abundance of the 0.1 eV gravitinos today will not overclose
the Universe. However, for the 1 keV gravitino, $\Omega_{\tilde G} \sim 1$.

 The decay products of gravitinos with zero temperature mass of 100 GeV will modify the light nuclei abundances adversely - the corresponding upper
bound on the gravitino abundance is $10^{-14}$ \cite{Cyburt:2009pg} 
which is 11 orders of magnitude
lower than the abundance obtained above. The gravitinos with zero temperature mass of 30 TeV will decay before nucleosynthesis and will not
modify the cosmology substantially.

Thus one needs to consider the cases of the 1 keV and 100 GeV mass gravitinos
carefully.  Now,
in the above analysis it was presumed that 
$\Treh >T_f$ which allowed the gravitinos to be in thermal equilibrium.
So to suppress the high abundance of the gravitinos as a hot relic one may consider
scenarios with $\Treh <T_f$ for the 1 keV and 100 GeV gravitinos.  
Then one has to consider out of equilibrium gravitino production till $T\sim m_0$
(when the scattering processes will be Boltzmann or kinematically suppressed)
using the Boltzmann 
equation, as in the standard calculation of the gravitino abundance.

We now consider the out of equilibrium production of gravitinos with a zero temperature mass of 100 GeV and $\Treh <T_f=1.2\times10^8\gev$, and with a zero temperature mass of 1 keV and
$\Treh <T_f=600\gev$.

\section{Out of equilibrium production of gravitinos}

 The gravitino production rate is given by the integrated Boltzmann equation
  \begin {equation}
  \frac{d{n}_{\tilde{G}}}{dt} +3 H n_{\tilde{G}}= \langle \Sigmatot |v|\rangle n^2. 
  \label{IntegBoltzEq}
  \end{equation} 
It is presumed that $n_{\tilde{G}} = 0$ at the beginning of the radiation dominated era 
after reheating and the gravitinos are
then produced through thermal scattering of the inflaton decay products.
We can rewrite Eq. \eqref{IntegBoltzEq} as
\begin{equation}
\dot T\frac{d Y_{\tilde{G}}}{dT}=n\langle \Sigmatot|v| \rangle
 Y = \Gamma_{s} Y
 \label{IntegBoltzEq1}
\end{equation}
where $Y={n}/{s}$ is the abundance of the scatterers. 
$T \propto \frac{1}{a}$, where $a$ is the scale factor of Universe. So
\bea
\frac{\dot T}{T}=-\frac{\dot a}{a}=-H=-\sqrt{\frac{8\pi G_N \rho}{3}}\\
= -\sqrt{\frac{8\pi G_N}{3}\frac{\pi^2}{30}g_* T^4}.
\eea
This gives
\begin{equation}\dot T=-\sqrt{\frac{g_*\pi^2}{90}}\frac{T^3}{M_P}.
\label{Tdot}
\end{equation}

Then,  on substituting Eqs. \eqref{Gammasfinal} and \eqref{Tdot} in Eq. \eqref{IntegBoltzEq1}, 
we obtain for spin 1/2 gravitinos
\be
\frac{d Y_{\tilde G}}{dT}=
-\frac{\beta\gamma_3}{\Mp}\,,
\label{YT-boltzmann}
\ee
where $\gamma_3$ is defined as in Eq. (\ref{susy1}) and
\bea
\beta&=& 
\left( \frac{90}{g_*\pi^2} \right)^{1/2}
\left(\frac{45}{2\pi^2 g_{*s}}\right) 
\left( \frac{\zeta(3)}{\pi^2} \right)^2\,\\
&&
\times\sum_{i=1}^3 \frac{3\pi}{16\zeta (3)}\,
c_i \,g_i^2\, \ln\left(\frac{k_i}{g_i}\right)\,.
\label{beta}
\eea
Hereafter we shall
assume 
$\beta$
to be independent of temperature and evaluate it at the dominant temperature limit in the integrals 
invoked below.

We are analyzing the case when $T_f>T_{\rm reh}\gg T$, which for $\mGz=100\gev$
can correspond to 
\textit {Regions II} or \textit{III}. Consider \textit{Region II} where $\delta_3 T^2>m_0^2$ 
and $\delta' T^2/(\sqrt{3}M_P)< m_{\tilde{G0}}$. Then
$\gamma_3={\delta_3 T^2}/(3 m_{\tilde G0}^2 )$ which gives  
\be
\frac{dY_{\tilde G}}{dT}\approx -\beta\frac{\delta_3 T^2}{3 \Mp m_{\tilde G0}^2}.
\label{boltzmann2a}
\ee 
On integrating  
from $\Treh$ to $T\ll\Treh$, we get
\bea
Y_{\tilde G}^{(1)}(T)&=& \beta\frac{\delta_3}{3 \Mp  m_{\tilde G0}^2}\frac{1}{3}\left(T_{\rm reh}^3-T^3\right)\\
&\approx & \beta\frac{\delta_3}{9 \Mp m_{\tilde G0}^2}T_{\rm reh}^3.
\label{Yhalf1}
\eea
Now consider \textit{Region III} when  $\delta_3 T^2<m_0^2$ and $\delta' T^2/(\sqrt{3}M_P)<m_{\tilde{G0}}$, i.e. $T<3m_0=300\gev$. Then 
$\gamma_3 = \frac{ m_0^2}{3m_{\tilde{G0}}^2} \approx \frac{1}{3}$,
as in the standard calculation.
On integrating Eq. \eqref{YT-boltzmann} from $T\sim3\,m_0$ to 
$T\sim m_0$, 
we get the
gravitino abundance 
\be
Y_\Gt^{(2)}= \frac{1}{3}\frac{\beta}{M_P} 2m_0.
 \ee
 Then the total gravitino abundance will be
 \bea
 Y_\Gt&=&Y_\Gt^{(1)}+Y_\Gt^{(2)}\cr
 &=&
\beta\frac{\delta_3}{9\Mp m_{\tilde G0}^2} T_{\rm reh}^3+ \frac{2}{3}\frac{\beta}{M_P} m_0\\
&\approx& \beta\frac{\delta_3}{9 \Mp m_{\tilde G0}^2} T_{\rm reh}^3,
\eea
as $Y_\Gt^{(2)}$ is much less than $Y_{\Gt}^{(1)}$. 

We find that the abundance of gravitinos is proportional to $T_{\rm reh}^3$. In order that the abundance lie within the cosmological bound of order $10^{-14} $ as given in Ref. \cite{Cyburt:2009pg}, $T_{\rm reh}$ should be less than $4\times10^4\gev$. 

For the 1 keV zero temperature mass gravitinos with $T_f=600\gev$, let $\Treh$ be 300 GeV.
This will correspond to \textit{Region III} with $\gamma_3 = \frac{ m_0^2}{3m_{\tilde{G0}}^2} =
3\times10^{15}$.  The gravitino abundance generated from $\Treh=300\gev$ to $T\sim m_0=100 \gev$ will be
\be
Y_\Gt= \gamma_3\frac{\beta}{M_P} 2m_0=1\times10^{-4}.
 \ee
Then from Eq. \eqref{OmegaG}
\be
\Omega_{\tilde G} = 0.05\,.
\ee
This is large and inconsistent with current observations.  This then implies that $\Treh$ must be
less than $m_0=100\gev$ to shut off this mode of gravitino production.

\section{Results and Conclusion}

By considering supersymmetry breaking due to the finite energy density of the Universe
we find that there is enhanced production of the spin 1/2 states of gravitinos 
(goldstino modes).  We have considered gravitinos with zero temperature masses of
0.1 eV, 1 keV, 100 GeV and 30 TeV as representative of gauge mediated (0.1 eV, 1 keV),
gravity mediated and anomaly mediated supersymmetry breaking scenarios respectively
and find that the production processes are in thermal equilibrium in the early Universe.
By studying the freeze out temperature for the gravitinos we have shown that
the gravitinos decouple as hot relics with large abundances.  In particular, the 1 keV and
100 GeV mass gravitinos have a very high abundance that can respectively close
the Universe or affect light nuclear abundances through their decay products.  

For both
these cases one can suppress the abundance by lowering the reheat temperature $\Treh$
below the freeze out temperature for gravitino production.  
Therefore we have further 
considered gravitino production from $\Treh$ below $T_f$ till $T\sim m_0 =100\gev$ (when the 
production shuts off)
using the Boltzmann equation.
For the 100 GeV gravitino with a freeze out temperature $T_f$ of $10^8\gev$, we
find that the abundance is proportional to $\Treh^3$ and that the
reheat  temperature must be less than $4\times10^4\gev$ to satisfy cosmological
constraints.  Such  a low reheat temperature will be inconsistent with  models of high scale baryogenesis
including those via leptogenesis.  Models of electroweak baryogenesis and low
scale leptogenesis \cite{Chun:2004eq,Boubekeur:2004ez,Dev:2015uca,Dev:2015cxa}
will then be preferred mechanisms for generating the 
matter-antimatter asymmetry of the Universe.

For the 1 keV gravitino with a freeze out temperature $T_f$ of $600\gev$, we first chose a 
reheat temperature of 300 GeV and calculated the abundance generated till $T\sim m_0$.
We found that the gravitinos will contribute 5\% of the total energy density of the Universe
today which is inconsistent with observations.  This implies that the reheat temperature should
be less than $m_0$.  Such a low reheat temperature may be obtained in models of electroweak scale inflation \cite{knoxturner} but will  rule out electroweak baryogenesis and leptogenesis
scenarios.
Then the preferred models of baryogenesis will be very low scale scenarios such as in 
Ref. \cite{Kohri:2009ka} or those involving neutron-antineutron oscillations \cite{Phillips:2014fgb} 
or black hole evaporation \cite{Majumdar:1995yr,
Nagatani:1998gv,Upadhyay:1999vk,Rangarajan:1999zp}.

The above analysis clearly provides a new manifestation of the gravitino problem.

\section*{Acknowledgements}
R. Rangarajan would like to thank Chandan Hati for useful discussions.

\end{document}